\documentclass[12pt]{article}
\usepackage{graphicx}
\usepackage{amssymb,amsmath}

\def\be{\begin{equation}}
\def\ee{\end{equation}}
\def\bc{\begin{center}}
\def\ec{\end{center}}
\def\bea{\begin{eqnarray}}
\def\eea{\end{eqnarray}}

\def\Kp00{\ensuremath{K^{+}\rightarrow\pi^{+}\pi^{0}\pi^{0}\,}}
\def\Kppm{\ensuremath{K^{+}\rightarrow\pi^{+}\pi^{+}\pi^{-}\,}}
\def\Kppp{\ensuremath{K^{+}\rightarrow\pi^{+}\pi^{+}\pi^{-}\,}}
\def\Kpzz{\ensuremath{K^{+}\rightarrow\pi^{+}\pi^{0}\pi^{0}\,}}
\def\Klzzz{\ensuremath{K_{L}\rightarrow\pi^{0}\pi^{0}\pi^{0}\,}}
\def\Klpmz{\ensuremath{K_{L}\rightarrow\pi^{+}\pi^{-}\pi^{0}\,}}
\def\p00{\ensuremath{\pi^{0}\pi^{0}\,}}
\def\pppm{\ensuremath{\pi^{+}\pi^{-}\,}}

\def\M{\ensuremath{\mathcal M}}
\def\mpp{\ensuremath{m_{\pi^{+}}}}
\def\spp{\ensuremath{s_{\pi\pi}\,}}
\def\ra{\ensuremath{\rightarrow\,}}

\newcommand{\av}{\operatorname{av}}
\newcommand{\cx}{\operatorname{chx}}
\newcommand{\thr}{\operatorname{thr}}

\catcode`@=11
\def\marginnote#1{}
\newcount\hour
\newcount\minute
\newtoks\amorpm
\hour=\time\divide\hour by60
\minute=\time{\multiply\hour by60 \global\advance\minute by-\hour}
\edef\standardtime{{\ifnum\hour<12 \global\amorpm={am}%
        \else\global\amorpm={pm}\advance\hour by-12 \fi
        \ifnum\hour=0 \hour=12 \fi
        \number\hour:\ifnum\minute<10 0\fi\number\minute\the\amorpm}}
\edef\militarytime{\number\hour:\ifnum\minute<10 0\fi\number\minute}
\def\draftlabel#1{{\@bsphack\if@filesw {\let\thepage\relax
   \xdef\@gtempa{\write\@auxout{\string
      \newlabel{#1}{{\@currentlabel}{\thepage}}}}}\@gtempa
   \if@nobreak \ifvmode\nobreak\fi\fi\fi\@esphack}
        \gdef\@eqnlabel{#1}}
\def\@eqnlabel{}
\def\@vacuum{}
\def\draftmarginnote#1{\marginpar{\raggedright\scriptsize\tt#1}}
\def\draft{\oddsidemargin 0.0truein
        \def\@oddfoot{\sl preliminary draft \hfil
        \rm\thepage\hfil\sl\today\quad\militarytime}
        \let\@evenfoot\@oddfoot \overfullrule 3pt
        \let\label=\draftlabel
        \let\marginnote=\draftmarginnote
   \def\@eqnnum{(\theequation)\rlap{\kern\marginparsep\tt\@eqnlabel}%
\global\let\@eqnlabel\@vacuum}  }
\catcode`@=12
\begin{document}
\begin{titlepage}
\vspace*{-1cm}
\hfill{CERN-PH-TH/2004-073}
\vskip 4.0cm
\bc
{\Large \bf Determination of the $a_{0}-a_{2}$ pion scattering length }
\ec
\bc
{\Large\bf from \Kp00 decay}
\ec
\vskip 1.0  cm
\begin{center}
{\large Nicola Cabibbo}~\footnote{e-mail address:
Nicola.cabibbo@roma1.infn.it}$^{\, ,}$~\footnote{On leave from
Universit\`a di Roma `La Sapienza' and INFN, Sezione di Roma, Italy}
\\
\vskip .1cm
CERN, Physics Department
\\ 
CH-1211 Geneva 23, Switzerland
\end{center}
\vskip 1.0cm
\begin{abstract}
\noindent
We present a new method for the determination of the $\pi -\pi$ scattering length
combination $a_{0}-a_{2}$, based on the study of the \p00 spectrum in \Kp00 in the
vicinity of the \pppm threshold. The method requires a minimum of theoretical input, and is
potentially very accurate. 
\end{abstract}
\end{titlepage}
\setcounter{footnote}{0}
\vskip2truecm
\setlength{\baselineskip}{.7cm}
%
%
Current algebra and PCAC lead to a prediction for the threshold behavior of $\pi -\pi$ scattering \cite{Weinberg:1966kf}\cite{Coleman}. The  $I=0$ and $I=2$ S-wave scattering lengths were predicted to be $a_{0}\mpp = 0.159,\, a_{2} \mpp= -0.045$, a first approximation that can be improved upon in the framework of Chiral Perturbation Theory \cite{ChPT}. Recent calculations \cite{Colangelo:2000jc}\cite{Colangelo01b}, which combine ChPT with the dispersive approach by S.~M.~Roy \cite{roy71}\cite{Ananthanarayan:2000ht}, lead to%
\be
\label{a02-1}
	a_{0}\mpp=0.220 \pm 0.005,\;\; a_{2}\mpp=-0.0444 \pm 0.0010,
	        \;\;(a_{0}- a_{2})\mpp=0.265\pm0.004
\ee
The current discussion  of this prediction, see 
 \cite{Pelaez:2003eh}\cite{Caprini:2003ta}\cite{Pelaez:2003ky}, could 
    lead to minor modifications of  eq. \eqref{a02-1}.
    
It was long recognized \cite{shabalin63} that the angular distributions in $K^{+}\rightarrow\pi^{+}\pi^{-}e^{+}\nu$ are sensitive to the $\pi\pi$ phase shifts, and can be used to obtain informations on the S-wave scattering lengths \cite{cabibbo65}\cite{pais67}. The first results by the Geneva-Saclay experiment \cite{rosselet77} , leading to $a_{0}\mpp=0.26\pm0.05$, where recently improved by the E865 experiment at Brookhaven \cite{Pislak:2003sv} that quotes a result: $a_{0}\mpp=0.216\pm0.013 \text{ (stat.)} \pm 0.002 \text{ (syst.)} \pm 0.002 \text{ (theor)} $. 
Data on $K_{e4}$, with a large statistics, are currently being analyzed by the NA48 experiment at CERN.

The $K_{e4}$ decay yields values of the phase shift difference $\delta^{0}_{0}-\delta^{1}_{1}$ as a function of the $\pi\pi$ invariant mass $\mu$ in the range $2\mpp<\mu<M_{K}-\mpp$,
but the best data lies in the range $>310$ MeV. The  extraction of a value for
$a_{0}$ requires an extrapolation to the threshold region and a substantial theoretical input, whence  the interest in alternative methods which permit the determination of the scattering lengths through measurements that are directly sensitive to $\pi\pi$ scattering in the threshold
region, $\mu\sim 2\mpp$. An example of this is the measurement of the $\pi^{0}\pi^{0}$ decay
of the pionic atom \pppm,  the object of the DIRAC experiment at CERN \cite{dirac} that could yield \cite{Gasser:1999vf} a value for the $a_{0}-a_{2}$ combination. 

I present here an alternative method for determining  $a_{0}-a_{2}$, based on
the $\pi^{0}\pi^{0}$ mass distribution in the \Kpzz decay in the vicinity of the \pppm  threshold. The  large data sample available from
the NA48 experiment at CERN, of the order of $10^{8}$ events, could lead to a determination of  $a_{0}-a_{2}$ with a precision comparable to that foreseen in the DIRAC experiment.
	\begin{figure}[h]
		\begin{center}
		\includegraphics[scale=0.90]{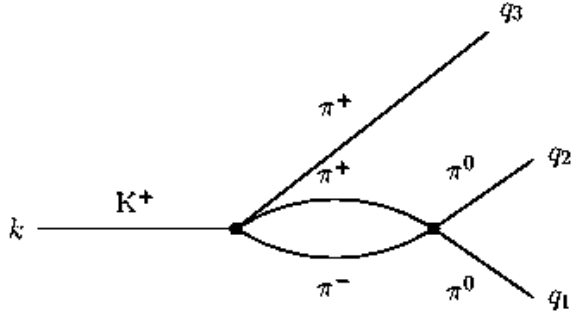}\\ 
		\end{center}
		\caption{The $\pi\pi$ re-scattering diagram. \label{fig1}}
	\end{figure}
The method is based on the fact that  the  \Kppm decay gives a contribution to the \Kp00 amplitude through the charge exchange reaction   $\pi^{+}\pi^{-} \rightarrow \pi^{0}\pi^{0}$. This contribution is directly proportional  to $a_{0}-a_{2}$, and displays a characteristic behavior when the $\pi^{0}\pi^{0}$ mass is in the vicinity of the \pppm threshold, where it goes from dispersive to (dominantly) absorptive.  

Let us write
\be
	\M(K^{+}\rightarrow\pi^{+}\pi^{0}\pi^{0}) = \M= \M_{0} + \M_{1}
\ee
where $\M_{0}$ is the ``unperturbed amplitude'', and $\M_{1}$ the contribution
of the diagram in Fig. \ref{fig1}, with the renormalization condition 
\be
	\M_{1}=0 \quad\text{for }\quad s_{\pi}=(q_{1}+q_{2})^{2}=4\mpp^{2}
\ee
The ``unperturbed'' amplitude $\M_{0}$, and the corresponding one $\M_{+}$ for
 \Kppp, can be parametrized as   polynomials \cite{PDG} in $s_{i}=(k-q_{i})^{2}$. In both cases $q_{3}$ is chosen as the momentum of the ``odd'' pion, respectively $\pi^{+}$ and $\pi^{-}$.  A simple parametrization,  which gives a reasonable description of the experimental data, is given by
 \begin{align}
 	\M_{0}(s_{1},s_{2},s_{3}) &= A_{\av}^{0}(1+g^{0}(s_{3}-s_{0})/2\mpp^{2})\label{Mzero}\\
 	\M_{+}(s_{1},s_{2},s_{3}) &= A_{\av}^{+}(1+g^{+}(s_{3}-s_{0})/2\mpp^{2})\label{Mplus},	
\end{align}
where  $s_{0}=(s_{1}+s_{2}+s_{3})/3$. The $g$'s coincide with the linear slope parameters defined in the PDG  review \cite{PDG}. The $\Delta I = 1/2$ rule  requires $A_{\av}^{0}$ and $A_{\av}^{+}$ to have the same sign \cite{sign}, and $A_{\av}^{+}\sim 2A_{\av}^{0}$ in good agreement with the observed branching ratios. In the following we will assume $\M_{0}$ and  $\M_{+}$ to be positive.

To evaluate the graph in Fig. \ref{fig1} we can use a simplified effective lagrangian which reproduces the $\pi\pi$ charge exchange reaction near the  \pppm threshold,
\be
  \label{rencon}
	\mathcal{L}_{\cx}= \frac{16\pi(a_{0}-a_{2})\mpp}{3}(\pi^{+}\pi^{-}\pi^{0}\pi^{0})
\ee
The diagram in Fig. \ref{fig1}  results then in: 
\be
   \label{Mres}
	\M_{1}= - \frac{2(a_{0}-a_{2})\mpp} {3}\M_{+,\thr}(J+K)
\ee
where $ \M_{+,\thr}$ is the value of $ \M_{+}$ at  the \pppm threshold. Using eq. \eqref{Mplus},
\be
\label{Mthr}
\M_{+,\thr}=A^{+}_{\av}\left(1+\frac{g^{+}(M_{K}^{2}-9\mpp^{2})}{12\mpp^{2}}\right)
\ee
We have divided the contribution of the graph into two parts, $J$ and $K$. The $J$ contribution flips from dispersive to absorptive at $s_{\pi\pi}=4\mpp^{2}$,
\be \label{J}
	\begin{array}{l@{\quad ;\quad} l@{\quad :\quad}l}
		J=J_{-}=\pi\tilde v&\tilde v=\sqrt{(4\mpp^{2}-s_{\pi\pi})/s_{\pi\pi}} & s_{\pi\pi}<4\mpp^{2}\\
		J=J_{+}=-i\pi v     & v= \sqrt{(s_{\pi\pi}-4\mpp^{2})/s_{\pi\pi}}         & s_{\pi\pi}>4\mpp^{2}
		\end{array} 
\ee
The K contribution is dispersive both above and below the threshold,
\be 
	\begin{array}{l@{\quad :\quad} l}
		K=-2 \tilde v  \arctan{\tilde v}= -2 \tilde v^{2} + \frac{2}{3}\tilde v^{4}+\ldots  & 
									s_{\pi\pi}<4\mpp^{2} \\
		K=-v \ln\left(\frac{1-v}{1+v}\right) =2v^{2}+ \frac{2}{3}v^{4}+\ldots & s_{\pi\pi}>4\mpp^{2} 
		\end{array} 
\ee
Noting that $\tilde v^{2}=-v^{2}$, the $K$ contribution can be expressed as a power series in $(s_{\pi\pi}-4\mpp^{2})$,  which converges when $|s_{\pi\pi}-4\mpp^{2}|<4\mpp^{2}$, a range which includes the physical region of $K_{3\pi}$ decays. This contribution can be approximated as a polynomial in \spp, so that we will reabsorb it in the definition of the ``unperturbed'' amplitude $\M_{0}$, setting $K=0$ in eq. \eqref{Mres}.

The differential decay rate  for \Kp00 with respect to the $\pi^{0}\pi^{0}$ invariant mass $M_{\pi\pi}=\sqrt{s_{\pi\pi}}$ is given by
\be
	\frac{d\Gamma}{dM_{\pi\pi}}=\left[\left(M_{\pi\pi}^{2}-4m_{\pi^{0}}^{2}\right)
						\left(1-\frac{(M_{\pi\pi}+m_{\pi^{+}})^{2}}{M_{K}^{2}}\right)
						\left(1-\frac{(M_{\pi\pi}-m_{\pi^{+}})^{2}}{M_{K}^{2}}\right)\right]^{\frac{1}{2}}
						 \left|\M\right |^{2}
\ee 
Since $\mathcal{M}^{1}$ changes from real to imaginary at the \pppm threshold,   we can write
\be
	\label{theres}
	\left|\M\right |^{2}= \left\{ \begin{array}{c@{\quad :\quad} l}
		(\M_{0})^{2}+(\M_{1})^{2}+2\M_{0}\M_{1} 	& s_{\pi\pi}<4\mpp^{2} \\
		(\M_{0})^{2}+(i\M_{1})^{2}				& s_{\pi\pi}>4\mpp^{2} 
		\end{array} \right . \\
\ee
	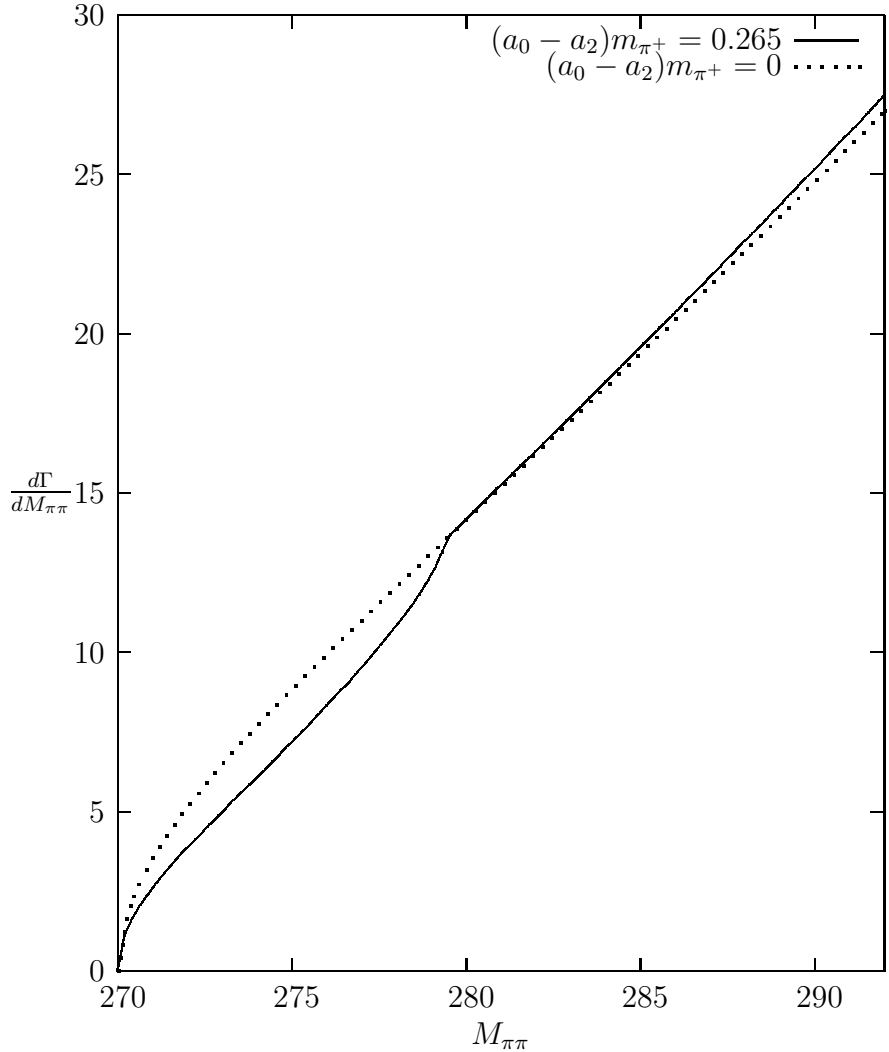
\begin{figure}[t]
		\begin{center}
\setlength{\unitlength}{0.240900pt}
\ifx\plotpoint\undefined\newsavebox{\plotpoint}\fi
\sbox{\plotpoint}{\rule[-0.200pt]{0.400pt}{0.400pt}}%
\begin{picture}(1425,1665)(0,0)
\sbox{\plotpoint}{\rule[-0.200pt]{0.400pt}{0.400pt}}%
\put(161.0,123.0){\rule[-0.200pt]{4.818pt}{0.400pt}}
\put(141,123){\makebox(0,0)[r]{ 0}}
\put(1344.0,123.0){\rule[-0.200pt]{4.818pt}{0.400pt}}
\put(161.0,373.0){\rule[-0.200pt]{4.818pt}{0.400pt}}
\put(141,373){\makebox(0,0)[r]{ 5}}
\put(1344.0,373.0){\rule[-0.200pt]{4.818pt}{0.400pt}}
\put(161.0,624.0){\rule[-0.200pt]{4.818pt}{0.400pt}}
\put(141,624){\makebox(0,0)[r]{ 10}}
\put(1344.0,624.0){\rule[-0.200pt]{4.818pt}{0.400pt}}
\put(161.0,874.0){\rule[-0.200pt]{4.818pt}{0.400pt}}
\put(141,874){\makebox(0,0)[r]{ 15}}
\put(1344.0,874.0){\rule[-0.200pt]{4.818pt}{0.400pt}}
\put(161.0,1124.0){\rule[-0.200pt]{4.818pt}{0.400pt}}
\put(141,1124){\makebox(0,0)[r]{ 20}}
\put(1344.0,1124.0){\rule[-0.200pt]{4.818pt}{0.400pt}}
\put(161.0,1375.0){\rule[-0.200pt]{4.818pt}{0.400pt}}
\put(141,1375){\makebox(0,0)[r]{ 25}}
\put(1344.0,1375.0){\rule[-0.200pt]{4.818pt}{0.400pt}}
\put(161.0,1625.0){\rule[-0.200pt]{4.818pt}{0.400pt}}
\put(141,1625){\makebox(0,0)[r]{ 30}}
\put(1344.0,1625.0){\rule[-0.200pt]{4.818pt}{0.400pt}}
\put(161.0,123.0){\rule[-0.200pt]{0.400pt}{4.818pt}}
\put(161,82){\makebox(0,0){ 270}}
\put(161.0,1605.0){\rule[-0.200pt]{0.400pt}{4.818pt}}
\put(434.0,123.0){\rule[-0.200pt]{0.400pt}{4.818pt}}
\put(434,82){\makebox(0,0){ 275}}
\put(434.0,1605.0){\rule[-0.200pt]{0.400pt}{4.818pt}}
\put(708.0,123.0){\rule[-0.200pt]{0.400pt}{4.818pt}}
\put(708,82){\makebox(0,0){ 280}}
\put(708.0,1605.0){\rule[-0.200pt]{0.400pt}{4.818pt}}
\put(981.0,123.0){\rule[-0.200pt]{0.400pt}{4.818pt}}
\put(981,82){\makebox(0,0){ 285}}
\put(981.0,1605.0){\rule[-0.200pt]{0.400pt}{4.818pt}}
\put(1255.0,123.0){\rule[-0.200pt]{0.400pt}{4.818pt}}
\put(1255,82){\makebox(0,0){ 290}}
\put(1255.0,1605.0){\rule[-0.200pt]{0.400pt}{4.818pt}}
\put(161.0,123.0){\rule[-0.200pt]{289.803pt}{0.400pt}}
\put(1364.0,123.0){\rule[-0.200pt]{0.400pt}{361.832pt}}
\put(161.0,1625.0){\rule[-0.200pt]{289.803pt}{0.400pt}}
\put(161.0,123.0){\rule[-0.200pt]{0.400pt}{361.832pt}}
\put(40,874){\makebox(0,0){$\frac{d\Gamma}{d M_{\pi\pi}}$}}
\put(762,21){\makebox(0,0){$M_{\pi\pi}$}}
\put(1204,1585){\makebox(0,0)[r]{$(a_0-a_2)m_{\pi^+}=0.265$}}
\put(1224.0,1585.0){\rule[-0.200pt]{24.090pt}{0.400pt}}
\put(161,123){\usebox{\plotpoint}}
\multiput(161.58,123.00)(0.492,2.521){21}{\rule{0.119pt}{2.067pt}}
\multiput(160.17,123.00)(12.000,54.711){2}{\rule{0.400pt}{1.033pt}}
\multiput(173.58,182.00)(0.492,1.056){21}{\rule{0.119pt}{0.933pt}}
\multiput(172.17,182.00)(12.000,23.063){2}{\rule{0.400pt}{0.467pt}}
\multiput(185.58,207.00)(0.492,0.841){21}{\rule{0.119pt}{0.767pt}}
\multiput(184.17,207.00)(12.000,18.409){2}{\rule{0.400pt}{0.383pt}}
\multiput(197.58,227.00)(0.493,0.695){23}{\rule{0.119pt}{0.654pt}}
\multiput(196.17,227.00)(13.000,16.643){2}{\rule{0.400pt}{0.327pt}}
\multiput(210.58,245.00)(0.492,0.669){21}{\rule{0.119pt}{0.633pt}}
\multiput(209.17,245.00)(12.000,14.685){2}{\rule{0.400pt}{0.317pt}}
\multiput(222.58,261.00)(0.492,0.625){21}{\rule{0.119pt}{0.600pt}}
\multiput(221.17,261.00)(12.000,13.755){2}{\rule{0.400pt}{0.300pt}}
\multiput(234.58,276.00)(0.492,0.582){21}{\rule{0.119pt}{0.567pt}}
\multiput(233.17,276.00)(12.000,12.824){2}{\rule{0.400pt}{0.283pt}}
\multiput(246.58,290.00)(0.492,0.582){21}{\rule{0.119pt}{0.567pt}}
\multiput(245.17,290.00)(12.000,12.824){2}{\rule{0.400pt}{0.283pt}}
\multiput(258.58,304.00)(0.492,0.539){21}{\rule{0.119pt}{0.533pt}}
\multiput(257.17,304.00)(12.000,11.893){2}{\rule{0.400pt}{0.267pt}}
\multiput(270.00,317.58)(0.497,0.493){23}{\rule{0.500pt}{0.119pt}}
\multiput(270.00,316.17)(11.962,13.000){2}{\rule{0.250pt}{0.400pt}}
\multiput(283.00,330.58)(0.496,0.492){21}{\rule{0.500pt}{0.119pt}}
\multiput(283.00,329.17)(10.962,12.000){2}{\rule{0.250pt}{0.400pt}}
\multiput(295.58,342.00)(0.492,0.539){21}{\rule{0.119pt}{0.533pt}}
\multiput(294.17,342.00)(12.000,11.893){2}{\rule{0.400pt}{0.267pt}}
\multiput(307.00,355.58)(0.496,0.492){21}{\rule{0.500pt}{0.119pt}}
\multiput(307.00,354.17)(10.962,12.000){2}{\rule{0.250pt}{0.400pt}}
\multiput(319.00,367.58)(0.496,0.492){21}{\rule{0.500pt}{0.119pt}}
\multiput(319.00,366.17)(10.962,12.000){2}{\rule{0.250pt}{0.400pt}}
\multiput(331.58,379.00)(0.492,0.539){21}{\rule{0.119pt}{0.533pt}}
\multiput(330.17,379.00)(12.000,11.893){2}{\rule{0.400pt}{0.267pt}}
\multiput(343.00,392.58)(0.496,0.492){21}{\rule{0.500pt}{0.119pt}}
\multiput(343.00,391.17)(10.962,12.000){2}{\rule{0.250pt}{0.400pt}}
\multiput(355.00,404.58)(0.539,0.492){21}{\rule{0.533pt}{0.119pt}}
\multiput(355.00,403.17)(11.893,12.000){2}{\rule{0.267pt}{0.400pt}}
\multiput(368.00,416.58)(0.496,0.492){21}{\rule{0.500pt}{0.119pt}}
\multiput(368.00,415.17)(10.962,12.000){2}{\rule{0.250pt}{0.400pt}}
\multiput(380.00,428.58)(0.496,0.492){21}{\rule{0.500pt}{0.119pt}}
\multiput(380.00,427.17)(10.962,12.000){2}{\rule{0.250pt}{0.400pt}}
\multiput(392.00,440.58)(0.496,0.492){21}{\rule{0.500pt}{0.119pt}}
\multiput(392.00,439.17)(10.962,12.000){2}{\rule{0.250pt}{0.400pt}}
\multiput(404.00,452.58)(0.496,0.492){21}{\rule{0.500pt}{0.119pt}}
\multiput(404.00,451.17)(10.962,12.000){2}{\rule{0.250pt}{0.400pt}}
\multiput(416.58,464.00)(0.492,0.539){21}{\rule{0.119pt}{0.533pt}}
\multiput(415.17,464.00)(12.000,11.893){2}{\rule{0.400pt}{0.267pt}}
\multiput(428.00,477.58)(0.496,0.492){21}{\rule{0.500pt}{0.119pt}}
\multiput(428.00,476.17)(10.962,12.000){2}{\rule{0.250pt}{0.400pt}}
\multiput(440.00,489.58)(0.497,0.493){23}{\rule{0.500pt}{0.119pt}}
\multiput(440.00,488.17)(11.962,13.000){2}{\rule{0.250pt}{0.400pt}}
\multiput(453.00,502.58)(0.496,0.492){21}{\rule{0.500pt}{0.119pt}}
\multiput(453.00,501.17)(10.962,12.000){2}{\rule{0.250pt}{0.400pt}}
\multiput(465.58,514.00)(0.492,0.539){21}{\rule{0.119pt}{0.533pt}}
\multiput(464.17,514.00)(12.000,11.893){2}{\rule{0.400pt}{0.267pt}}
\multiput(477.58,527.00)(0.492,0.539){21}{\rule{0.119pt}{0.533pt}}
\multiput(476.17,527.00)(12.000,11.893){2}{\rule{0.400pt}{0.267pt}}
\multiput(489.58,540.00)(0.492,0.539){21}{\rule{0.119pt}{0.533pt}}
\multiput(488.17,540.00)(12.000,11.893){2}{\rule{0.400pt}{0.267pt}}
\multiput(501.58,553.00)(0.492,0.539){21}{\rule{0.119pt}{0.533pt}}
\multiput(500.17,553.00)(12.000,11.893){2}{\rule{0.400pt}{0.267pt}}
\multiput(513.00,566.58)(0.497,0.493){23}{\rule{0.500pt}{0.119pt}}
\multiput(513.00,565.17)(11.962,13.000){2}{\rule{0.250pt}{0.400pt}}
\multiput(526.58,579.00)(0.492,0.582){21}{\rule{0.119pt}{0.567pt}}
\multiput(525.17,579.00)(12.000,12.824){2}{\rule{0.400pt}{0.283pt}}
\multiput(538.58,593.00)(0.492,0.539){21}{\rule{0.119pt}{0.533pt}}
\multiput(537.17,593.00)(12.000,11.893){2}{\rule{0.400pt}{0.267pt}}
\multiput(550.58,606.00)(0.492,0.625){21}{\rule{0.119pt}{0.600pt}}
\multiput(549.17,606.00)(12.000,13.755){2}{\rule{0.400pt}{0.300pt}}
\multiput(562.58,621.00)(0.492,0.582){21}{\rule{0.119pt}{0.567pt}}
\multiput(561.17,621.00)(12.000,12.824){2}{\rule{0.400pt}{0.283pt}}
\multiput(574.58,635.00)(0.492,0.625){21}{\rule{0.119pt}{0.600pt}}
\multiput(573.17,635.00)(12.000,13.755){2}{\rule{0.400pt}{0.300pt}}
\multiput(586.58,650.00)(0.492,0.625){21}{\rule{0.119pt}{0.600pt}}
\multiput(585.17,650.00)(12.000,13.755){2}{\rule{0.400pt}{0.300pt}}
\multiput(598.58,665.00)(0.493,0.616){23}{\rule{0.119pt}{0.592pt}}
\multiput(597.17,665.00)(13.000,14.771){2}{\rule{0.400pt}{0.296pt}}
\multiput(611.58,681.00)(0.492,0.669){21}{\rule{0.119pt}{0.633pt}}
\multiput(610.17,681.00)(12.000,14.685){2}{\rule{0.400pt}{0.317pt}}
\multiput(623.58,697.00)(0.492,0.755){21}{\rule{0.119pt}{0.700pt}}
\multiput(622.17,697.00)(12.000,16.547){2}{\rule{0.400pt}{0.350pt}}
\multiput(635.58,715.00)(0.492,0.798){21}{\rule{0.119pt}{0.733pt}}
\multiput(634.17,715.00)(12.000,17.478){2}{\rule{0.400pt}{0.367pt}}
\multiput(647.58,734.00)(0.492,0.884){21}{\rule{0.119pt}{0.800pt}}
\multiput(646.17,734.00)(12.000,19.340){2}{\rule{0.400pt}{0.400pt}}
\multiput(659.58,755.00)(0.492,1.142){21}{\rule{0.119pt}{1.000pt}}
\multiput(658.17,755.00)(12.000,24.924){2}{\rule{0.400pt}{0.500pt}}
\multiput(671.58,782.00)(0.493,1.091){23}{\rule{0.119pt}{0.962pt}}
\multiput(670.17,782.00)(13.000,26.004){2}{\rule{0.400pt}{0.481pt}}
\multiput(684.00,810.58)(0.543,0.492){19}{\rule{0.536pt}{0.118pt}}
\multiput(684.00,809.17)(10.887,11.000){2}{\rule{0.268pt}{0.400pt}}
\multiput(696.00,821.58)(0.496,0.492){21}{\rule{0.500pt}{0.119pt}}
\multiput(696.00,820.17)(10.962,12.000){2}{\rule{0.250pt}{0.400pt}}
\multiput(708.00,833.58)(0.496,0.492){21}{\rule{0.500pt}{0.119pt}}
\multiput(708.00,832.17)(10.962,12.000){2}{\rule{0.250pt}{0.400pt}}
\multiput(720.00,845.58)(0.496,0.492){21}{\rule{0.500pt}{0.119pt}}
\multiput(720.00,844.17)(10.962,12.000){2}{\rule{0.250pt}{0.400pt}}
\multiput(732.00,857.58)(0.496,0.492){21}{\rule{0.500pt}{0.119pt}}
\multiput(732.00,856.17)(10.962,12.000){2}{\rule{0.250pt}{0.400pt}}
\multiput(744.00,869.58)(0.496,0.492){21}{\rule{0.500pt}{0.119pt}}
\multiput(744.00,868.17)(10.962,12.000){2}{\rule{0.250pt}{0.400pt}}
\multiput(756.00,881.58)(0.539,0.492){21}{\rule{0.533pt}{0.119pt}}
\multiput(756.00,880.17)(11.893,12.000){2}{\rule{0.267pt}{0.400pt}}
\multiput(769.00,893.58)(0.496,0.492){21}{\rule{0.500pt}{0.119pt}}
\multiput(769.00,892.17)(10.962,12.000){2}{\rule{0.250pt}{0.400pt}}
\multiput(781.00,905.58)(0.496,0.492){21}{\rule{0.500pt}{0.119pt}}
\multiput(781.00,904.17)(10.962,12.000){2}{\rule{0.250pt}{0.400pt}}
\multiput(793.00,917.58)(0.543,0.492){19}{\rule{0.536pt}{0.118pt}}
\multiput(793.00,916.17)(10.887,11.000){2}{\rule{0.268pt}{0.400pt}}
\multiput(805.00,928.58)(0.496,0.492){21}{\rule{0.500pt}{0.119pt}}
\multiput(805.00,927.17)(10.962,12.000){2}{\rule{0.250pt}{0.400pt}}
\multiput(817.00,940.58)(0.496,0.492){21}{\rule{0.500pt}{0.119pt}}
\multiput(817.00,939.17)(10.962,12.000){2}{\rule{0.250pt}{0.400pt}}
\multiput(829.00,952.58)(0.496,0.492){21}{\rule{0.500pt}{0.119pt}}
\multiput(829.00,951.17)(10.962,12.000){2}{\rule{0.250pt}{0.400pt}}
\multiput(841.00,964.58)(0.539,0.492){21}{\rule{0.533pt}{0.119pt}}
\multiput(841.00,963.17)(11.893,12.000){2}{\rule{0.267pt}{0.400pt}}
\multiput(854.00,976.58)(0.496,0.492){21}{\rule{0.500pt}{0.119pt}}
\multiput(854.00,975.17)(10.962,12.000){2}{\rule{0.250pt}{0.400pt}}
\multiput(866.00,988.58)(0.496,0.492){21}{\rule{0.500pt}{0.119pt}}
\multiput(866.00,987.17)(10.962,12.000){2}{\rule{0.250pt}{0.400pt}}
\multiput(878.00,1000.58)(0.496,0.492){21}{\rule{0.500pt}{0.119pt}}
\multiput(878.00,999.17)(10.962,12.000){2}{\rule{0.250pt}{0.400pt}}
\multiput(890.00,1012.58)(0.496,0.492){21}{\rule{0.500pt}{0.119pt}}
\multiput(890.00,1011.17)(10.962,12.000){2}{\rule{0.250pt}{0.400pt}}
\multiput(902.00,1024.58)(0.496,0.492){21}{\rule{0.500pt}{0.119pt}}
\multiput(902.00,1023.17)(10.962,12.000){2}{\rule{0.250pt}{0.400pt}}
\multiput(914.00,1036.58)(0.497,0.493){23}{\rule{0.500pt}{0.119pt}}
\multiput(914.00,1035.17)(11.962,13.000){2}{\rule{0.250pt}{0.400pt}}
\multiput(927.00,1049.58)(0.496,0.492){21}{\rule{0.500pt}{0.119pt}}
\multiput(927.00,1048.17)(10.962,12.000){2}{\rule{0.250pt}{0.400pt}}
\multiput(939.00,1061.58)(0.496,0.492){21}{\rule{0.500pt}{0.119pt}}
\multiput(939.00,1060.17)(10.962,12.000){2}{\rule{0.250pt}{0.400pt}}
\multiput(951.00,1073.58)(0.496,0.492){21}{\rule{0.500pt}{0.119pt}}
\multiput(951.00,1072.17)(10.962,12.000){2}{\rule{0.250pt}{0.400pt}}
\multiput(963.00,1085.58)(0.496,0.492){21}{\rule{0.500pt}{0.119pt}}
\multiput(963.00,1084.17)(10.962,12.000){2}{\rule{0.250pt}{0.400pt}}
\multiput(975.00,1097.58)(0.496,0.492){21}{\rule{0.500pt}{0.119pt}}
\multiput(975.00,1096.17)(10.962,12.000){2}{\rule{0.250pt}{0.400pt}}
\multiput(987.00,1109.58)(0.496,0.492){21}{\rule{0.500pt}{0.119pt}}
\multiput(987.00,1108.17)(10.962,12.000){2}{\rule{0.250pt}{0.400pt}}
\multiput(999.00,1121.58)(0.497,0.493){23}{\rule{0.500pt}{0.119pt}}
\multiput(999.00,1120.17)(11.962,13.000){2}{\rule{0.250pt}{0.400pt}}
\multiput(1012.00,1134.58)(0.496,0.492){21}{\rule{0.500pt}{0.119pt}}
\multiput(1012.00,1133.17)(10.962,12.000){2}{\rule{0.250pt}{0.400pt}}
\multiput(1024.00,1146.58)(0.496,0.492){21}{\rule{0.500pt}{0.119pt}}
\multiput(1024.00,1145.17)(10.962,12.000){2}{\rule{0.250pt}{0.400pt}}
\multiput(1036.00,1158.58)(0.496,0.492){21}{\rule{0.500pt}{0.119pt}}
\multiput(1036.00,1157.17)(10.962,12.000){2}{\rule{0.250pt}{0.400pt}}
\multiput(1048.58,1170.00)(0.492,0.539){21}{\rule{0.119pt}{0.533pt}}
\multiput(1047.17,1170.00)(12.000,11.893){2}{\rule{0.400pt}{0.267pt}}
\multiput(1060.00,1183.58)(0.496,0.492){21}{\rule{0.500pt}{0.119pt}}
\multiput(1060.00,1182.17)(10.962,12.000){2}{\rule{0.250pt}{0.400pt}}
\multiput(1072.00,1195.58)(0.539,0.492){21}{\rule{0.533pt}{0.119pt}}
\multiput(1072.00,1194.17)(11.893,12.000){2}{\rule{0.267pt}{0.400pt}}
\multiput(1085.58,1207.00)(0.492,0.539){21}{\rule{0.119pt}{0.533pt}}
\multiput(1084.17,1207.00)(12.000,11.893){2}{\rule{0.400pt}{0.267pt}}
\multiput(1097.00,1220.58)(0.496,0.492){21}{\rule{0.500pt}{0.119pt}}
\multiput(1097.00,1219.17)(10.962,12.000){2}{\rule{0.250pt}{0.400pt}}
\multiput(1109.58,1232.00)(0.492,0.539){21}{\rule{0.119pt}{0.533pt}}
\multiput(1108.17,1232.00)(12.000,11.893){2}{\rule{0.400pt}{0.267pt}}
\multiput(1121.00,1245.58)(0.496,0.492){21}{\rule{0.500pt}{0.119pt}}
\multiput(1121.00,1244.17)(10.962,12.000){2}{\rule{0.250pt}{0.400pt}}
\multiput(1133.58,1257.00)(0.492,0.539){21}{\rule{0.119pt}{0.533pt}}
\multiput(1132.17,1257.00)(12.000,11.893){2}{\rule{0.400pt}{0.267pt}}
\multiput(1145.00,1270.58)(0.496,0.492){21}{\rule{0.500pt}{0.119pt}}
\multiput(1145.00,1269.17)(10.962,12.000){2}{\rule{0.250pt}{0.400pt}}
\multiput(1157.00,1282.58)(0.497,0.493){23}{\rule{0.500pt}{0.119pt}}
\multiput(1157.00,1281.17)(11.962,13.000){2}{\rule{0.250pt}{0.400pt}}
\multiput(1170.00,1295.58)(0.496,0.492){21}{\rule{0.500pt}{0.119pt}}
\multiput(1170.00,1294.17)(10.962,12.000){2}{\rule{0.250pt}{0.400pt}}
\multiput(1182.58,1307.00)(0.492,0.539){21}{\rule{0.119pt}{0.533pt}}
\multiput(1181.17,1307.00)(12.000,11.893){2}{\rule{0.400pt}{0.267pt}}
\multiput(1194.00,1320.58)(0.496,0.492){21}{\rule{0.500pt}{0.119pt}}
\multiput(1194.00,1319.17)(10.962,12.000){2}{\rule{0.250pt}{0.400pt}}
\multiput(1206.58,1332.00)(0.492,0.539){21}{\rule{0.119pt}{0.533pt}}
\multiput(1205.17,1332.00)(12.000,11.893){2}{\rule{0.400pt}{0.267pt}}
\multiput(1218.00,1345.58)(0.496,0.492){21}{\rule{0.500pt}{0.119pt}}
\multiput(1218.00,1344.17)(10.962,12.000){2}{\rule{0.250pt}{0.400pt}}
\multiput(1230.58,1357.00)(0.492,0.539){21}{\rule{0.119pt}{0.533pt}}
\multiput(1229.17,1357.00)(12.000,11.893){2}{\rule{0.400pt}{0.267pt}}
\multiput(1242.00,1370.58)(0.497,0.493){23}{\rule{0.500pt}{0.119pt}}
\multiput(1242.00,1369.17)(11.962,13.000){2}{\rule{0.250pt}{0.400pt}}
\multiput(1255.00,1383.58)(0.496,0.492){21}{\rule{0.500pt}{0.119pt}}
\multiput(1255.00,1382.17)(10.962,12.000){2}{\rule{0.250pt}{0.400pt}}
\multiput(1267.58,1395.00)(0.492,0.539){21}{\rule{0.119pt}{0.533pt}}
\multiput(1266.17,1395.00)(12.000,11.893){2}{\rule{0.400pt}{0.267pt}}
\multiput(1279.58,1408.00)(0.492,0.539){21}{\rule{0.119pt}{0.533pt}}
\multiput(1278.17,1408.00)(12.000,11.893){2}{\rule{0.400pt}{0.267pt}}
\multiput(1291.58,1421.00)(0.492,0.539){21}{\rule{0.119pt}{0.533pt}}
\multiput(1290.17,1421.00)(12.000,11.893){2}{\rule{0.400pt}{0.267pt}}
\multiput(1303.00,1434.58)(0.496,0.492){21}{\rule{0.500pt}{0.119pt}}
\multiput(1303.00,1433.17)(10.962,12.000){2}{\rule{0.250pt}{0.400pt}}
\multiput(1315.00,1446.58)(0.497,0.493){23}{\rule{0.500pt}{0.119pt}}
\multiput(1315.00,1445.17)(11.962,13.000){2}{\rule{0.250pt}{0.400pt}}
\multiput(1328.58,1459.00)(0.492,0.539){21}{\rule{0.119pt}{0.533pt}}
\multiput(1327.17,1459.00)(12.000,11.893){2}{\rule{0.400pt}{0.267pt}}
\multiput(1340.58,1472.00)(0.492,0.539){21}{\rule{0.119pt}{0.533pt}}
\multiput(1339.17,1472.00)(12.000,11.893){2}{\rule{0.400pt}{0.267pt}}
\multiput(1352.58,1485.00)(0.492,0.539){21}{\rule{0.119pt}{0.533pt}}
\multiput(1351.17,1485.00)(12.000,11.893){2}{\rule{0.400pt}{0.267pt}}
\sbox{\plotpoint}{\rule[-0.500pt]{1.000pt}{1.000pt}}%
\sbox{\plotpoint}{\rule[-0.200pt]{0.400pt}{0.400pt}}%
\put(1204,1544){\makebox(0,0)[r]{$(a_0-a_2)m_{\pi^+}=0$}}
\sbox{\plotpoint}{\rule[-0.500pt]{1.000pt}{1.000pt}}%
\multiput(1224,1544)(20.756,0.000){5}{\usebox{\plotpoint}}
\put(1324,1544){\usebox{\plotpoint}}
\put(161,123){\usebox{\plotpoint}}
\multiput(161,123)(3.005,20.537){4}{\usebox{\plotpoint}}
\multiput(173,205)(6.732,19.634){2}{\usebox{\plotpoint}}
\multiput(185,240)(8.176,19.077){2}{\usebox{\plotpoint}}
\put(204.73,281.67){\usebox{\plotpoint}}
\put(214.98,299.72){\usebox{\plotpoint}}
\put(225.53,317.59){\usebox{\plotpoint}}
\put(236.72,335.07){\usebox{\plotpoint}}
\put(248.32,352.28){\usebox{\plotpoint}}
\put(260.29,369.24){\usebox{\plotpoint}}
\put(272.56,385.96){\usebox{\plotpoint}}
\put(285.89,401.85){\usebox{\plotpoint}}
\put(298.63,418.23){\usebox{\plotpoint}}
\put(311.93,434.16){\usebox{\plotpoint}}
\put(325.14,450.16){\usebox{\plotpoint}}
\put(338.97,465.64){\usebox{\plotpoint}}
\put(352.64,481.25){\usebox{\plotpoint}}
\put(367.11,496.11){\usebox{\plotpoint}}
\put(381.23,511.33){\usebox{\plotpoint}}
\put(395.31,526.58){\usebox{\plotpoint}}
\put(409.61,541.61){\usebox{\plotpoint}}
\put(423.95,556.61){\usebox{\plotpoint}}
\put(438.46,571.46){\usebox{\plotpoint}}
\put(453.13,586.13){\usebox{\plotpoint}}
\put(467.81,600.81){\usebox{\plotpoint}}
\put(482.49,615.48){\usebox{\plotpoint}}
\put(497.16,630.16){\usebox{\plotpoint}}
\put(511.84,644.84){\usebox{\plotpoint}}
\put(527.00,659.00){\usebox{\plotpoint}}
\put(541.84,673.52){\usebox{\plotpoint}}
\put(556.85,687.85){\usebox{\plotpoint}}
\put(571.52,702.52){\usebox{\plotpoint}}
\put(586.21,717.19){\usebox{\plotpoint}}
\put(601.50,731.23){\usebox{\plotpoint}}
\put(616.77,745.29){\usebox{\plotpoint}}
\put(631.70,759.70){\usebox{\plotpoint}}
\put(646.37,774.37){\usebox{\plotpoint}}
\put(661.54,788.54){\usebox{\plotpoint}}
\put(676.63,802.76){\usebox{\plotpoint}}
\put(691.85,816.85){\usebox{\plotpoint}}
\put(706.53,831.53){\usebox{\plotpoint}}
\put(721.69,845.69){\usebox{\plotpoint}}
\put(736.55,860.17){\usebox{\plotpoint}}
\put(751.53,874.53){\usebox{\plotpoint}}
\put(767.02,888.33){\usebox{\plotpoint}}
\put(781.88,902.81){\usebox{\plotpoint}}
\put(797.01,917.01){\usebox{\plotpoint}}
\put(811.69,931.69){\usebox{\plotpoint}}
\put(826.76,945.95){\usebox{\plotpoint}}
\put(841.57,960.48){\usebox{\plotpoint}}
\put(857.16,974.16){\usebox{\plotpoint}}
\put(871.84,988.84){\usebox{\plotpoint}}
\put(886.88,1003.14){\usebox{\plotpoint}}
\put(901.68,1017.68){\usebox{\plotpoint}}
\put(916.54,1032.15){\usebox{\plotpoint}}
\put(931.99,1045.99){\usebox{\plotpoint}}
\put(946.67,1060.67){\usebox{\plotpoint}}
\put(961.34,1075.34){\usebox{\plotpoint}}
\put(976.51,1089.51){\usebox{\plotpoint}}
\put(991.19,1104.19){\usebox{\plotpoint}}
\put(1006.13,1118.58){\usebox{\plotpoint}}
\put(1021.41,1132.63){\usebox{\plotpoint}}
\put(1036.19,1147.19){\usebox{\plotpoint}}
\put(1050.87,1161.87){\usebox{\plotpoint}}
\put(1065.55,1176.55){\usebox{\plotpoint}}
\put(1080.55,1190.89){\usebox{\plotpoint}}
\put(1095.39,1205.39){\usebox{\plotpoint}}
\put(1110.07,1220.07){\usebox{\plotpoint}}
\put(1124.74,1234.74){\usebox{\plotpoint}}
\put(1139.42,1249.42){\usebox{\plotpoint}}
\put(1154.10,1264.10){\usebox{\plotpoint}}
\put(1169.23,1278.29){\usebox{\plotpoint}}
\put(1183.94,1292.94){\usebox{\plotpoint}}
\put(1198.61,1307.61){\usebox{\plotpoint}}
\put(1213.29,1322.29){\usebox{\plotpoint}}
\put(1227.97,1336.97){\usebox{\plotpoint}}
\put(1242.14,1352.13){\usebox{\plotpoint}}
\put(1257.30,1366.30){\usebox{\plotpoint}}
\put(1271.98,1380.98){\usebox{\plotpoint}}
\put(1286.34,1395.95){\usebox{\plotpoint}}
\put(1300.82,1410.82){\usebox{\plotpoint}}
\put(1315.51,1425.47){\usebox{\plotpoint}}
\put(1330.55,1439.77){\usebox{\plotpoint}}
\put(1344.83,1454.83){\usebox{\plotpoint}}
\put(1359.50,1469.50){\usebox{\plotpoint}}
\put(1364,1474){\usebox{\plotpoint}}
\sbox{\plotpoint}{\rule[-0.200pt]{0.400pt}{0.400pt}}%
\put(161.0,123.0){\rule[-0.200pt]{289.803pt}{0.400pt}}
\put(1364.0,123.0){\rule[-0.200pt]{0.400pt}{361.832pt}}
\put(161.0,1625.0){\rule[-0.200pt]{289.803pt}{0.400pt}}
\put(161.0,123.0){\rule[-0.200pt]{0.400pt}{361.832pt}}
\end{picture}
		\end{center}
		\caption{The \p00 invariant mass distribution 
		with/without the re-scattering correction, in arbitrary units.}
		\label{fig2}
	\end{figure}
In Fig.\ref{fig2} we show a plot of the differential decay rate (in arbitrary units) before and after the re-scattering corrections, evaluated using $A_{\av}^{+}=2A_{\av}^{0}$, the slope parameters $g^{\pm}$ as given in the PDG listings, and the value for $a_{0}-a_{2}$ from eq. \eqref{a02-1}.  The $\sqrt{4\mpp^{2}-s_{\pi\pi}}$ behavior below the \pppm threshold arises from the interference term in eq. \eqref{theres} and is a very characteristic feature.
It is encouraging to see that the deviation from the uncorrected behavior is very prominent, so that it should be possible to measure it accurately.

In order to extract the value of $a_{0}-a_{2}$ from the \p00 spectrum, let us consider a development
of $|\M|^{2}$ in powers of $\delta=\sqrt{(4\mpp^{2}-s_{\pi\pi})/4\mpp^{2}}$. Below the \pppm threshold the  coefficients of  $ \delta$ and of $ \delta^{2}$ are uniquely determined in terms of the rate for \Kp00 above this threshold,  the \Kppm differential rate, and the value of $a_{0}-a_{2}$. Since 
the maximum value of $\delta$  below threshold is  $\sim 0.26$, neglecting terms in $\delta^{3}$ and higher is equivalent to a theoretical error of $\sim 2\%$. This is the central result of this paper, 
and it is worthwhile to discuss  it in more detail.

Above the \pppm threshold $\M_{1}$ is absorptive, so that its value is directly determined by the
physical amplitudes for \Kppm and $\pppm\rightarrow\p00$ (eqs. \ref{Mres}, \ref{J}).   In eqs. \eqref{Mres}, \eqref{J} we have neglected the \spp dependence of the charge exchange reaction and of the \Kppm rate, which can contribute terms of O($\delta^{3}$) to  $\M_{1}$. 
As noted before in the discussion of the $K$ term, even powers of $\delta$ are absent from $\M_{1}$ because they can be absorbed in the definition of $\M_{0}$. The value of $\M_{1}$ below the threshold is the analytic continuation of the value above the threshold, so that it correctly includes the O($\delta$) terms, with possible errors which are O($\delta^{3}$).

Terms of   O($\delta^{2}=(4\mpp^{2}-s_{\pi\pi})/4\mpp^{2}$) in the value of  $|\M|^{2}$, eq. \eqref{theres}, derive from two sources: the first
is in the \spp dependence of $\M_{0}$ --- see e.g. eq. \eqref{Mzero}, keeping in mind that $s_{3}=(k-q_{3})^{2}=(q_{1}+q_{2})^{2}=s_{\pi\pi}$. Since $\M_{0}$ is regular at the threshold, the coefficient of this contribution is the same on either  side of it. The second source of  O($\delta^{2}$) terms is from the
$(\M_{1})^{2}$ terms in eq. \eqref{theres}. In this case, since $\tilde v^{2}=-v^{2}$, the coefficient of
$\delta^{2}$ changes sign across the threshold. This coefficient is predicted by eqs. \eqref{Mres}, \eqref{J}. We can thus proceed as follows:
\begin{enumerate}
\item Measure $\M_{+,\thr}$ from the \Kppm decay at the \pppm threshold. In terms of the PDG inspired parametrization in eq. \eqref{Mplus}, $\M_{+,\thr}$  is given by eq. \eqref{Mthr}.
\item Fit $|\M|^{2} = (\M_{0})^{2}+(i\M_{1})^{2} $, measured  from \Kp00 with $M_{\pi\pi}$ above the \pppm threshold, to  a polynomial in $\delta^{2}$, $|\M|^{2}=F(\delta^{2})$. 
\item  $|\M|^{2}$ below the threshold will then  be given by 
\be
\label{Mfit}
	|\M|^{2} = F(\delta^{2}) +2\M_{1}\sqrt{F(\delta^{2})+(\M_{1})^{2}} + 2(\M_{1} )^{2}
\ee
  where $F(\delta^{2})$ is the polynomial obtained in the second step.
\item Using  eqs. \eqref{Mres}, \eqref{J}, we can express $\M_{1}$ in terms of $a_{0}-a_{2}$, so that this quantity can be obtained by fitting the \p00 spectrum below the \pppm threshold to eq. \eqref{Mfit}. 
\end{enumerate}
We have not so far discussed the contribution $\M_{2}$ of the diagram, similar to 
  that in Fig.~\ref{fig1}, which arises from the unperturbed amplitude $\M_{0}$ with \p00\ra\p00 
  re-scattering. This contribution is always absorptive, and generally smaller than  $\M_{1}$. 
  It does not interfere with $\M_{0}$, but it interferes with $\M_{1}$ above the \pppm threshold. 
The effects of  $\M_{2}$ are small and will not impact on the precision of $a_{0}-a_{2}$, but should be included in the analysis of the experimental data,  with a slight complication of the fitting procedure we have outlined. For completeness we register its value \cite{whichpi}:
\be
	\M_{2}= - \frac{(a_{0}+2a_{2})m_{\pi^{0}}} {3}\M_{0,\thr}(-i \sqrt{1-\frac{4m_{\pi^{0}}^{2}}{s_{\pi\pi}}})
\ee
where $\M_{0,\thr}$ is the unperturbed amplitude at the \p00 threshold. Since the effects of this amplitude are small, the experiment will not be very sensitive to $(a_{0}+2a_{2})$,
and the best strategy could be to accept for it the theoretical prediction from eq. \eqref{a02-1}, while
extracting a value for $(a_{0}-a_{2})$.
  
  Although the method outlined here seems to require a minimum of theoretical elaboration, more theoretical work is needed. Given the possible precision of the method, it would be nice to obtain a more exact evaluation of the O($\delta^{3}$) corrections to $|\M|^{2}$. This will be possible with the methods of Chiral Perturbation Theory. It is of course possible to account for these corrections by introducing an extra parameter in the fit to the experimental data. We might also wish  to evaluate the electromagnetic corrections to our predictions.

We note that a similar effect arises in the interference between \Klzzz and
\Klpmz followed by \pppm \ra \p00. The effect is smaller than in fig. \ref{fig2},
but could also lead to a determination of $a_{0}-a_{2}$. Similar effects should also appear in 
$\eta\ra 3\pi^{0}$ decays, but this process is not competitive from an experimental point of view.

Threshold cusp phenomena have a long history \cite{Wigner}\cite{Breit}. They have been studied in $\pi^{-}P\ra\Lambda K^{0}$ near the $\Sigma K$ threshold \cite{Adair}\cite{Nelson} in an attempt to determine the relative $\Sigma-\Lambda$ parity, and more recently \cite{Bernstein:1996vf} in
$\gamma P\ra \pi^{0} P$ near the $N\pi^{+}$ threshold, where they can yield informations on the 
$\pi$--nucleon scattering lengths. In contrast to the phenomenon discussed here, the analysis of
cusp phenomena in two-body processes is inherently more complex.
\vspace{5mm}

I am grateful to Italo Mannelli and to  Augusto Ceccucci for discussions of the early results on the \p00 spectrum which inspired the present work, and to Roland Winston for a discussion of the early history of threshold cusps.


\begin{thebibliography}{99}
\bibitem{sign}
A verification of this fact would be one of the results of the proposed measurement.
\bibitem{whichpi}
Between the  \p00 and the \pppm thresholds I-spin is clearly broken. Here we are implicitly defining $(a_{0}+2a_{2})/3$ as the scattering length which describes \p00\ra\p00 at the \p00 threshold, and eq. \eqref{Mres} defines  $(a_{0}-a_{2})/3$ as the scattering length  which describes
\pppm\ra\p00 at the \pppm threshold --- the factor 2 in eq. \eqref{Mres}  arising from the two \pppm combinations from \Kppm which can re-scatter into \p00. The relationship between these quantities and the scattering lengths measured far from the thresholds would be more exactly determined by radiative corrections.
\end{thebibliography}
\end{document}